\documentclass[twocolumn]{aastex62}
\usepackage{CJKutf8}
 
\newcommand{\OII}{[\ion{O}{2}]} 
\newcommand{\OIII}{[\ion{O}{3}]} 
\newcommand{\NII}{[\ion{N}{2}]}
\newcommand{\SII}{[\ion{S}{2}]}  
\newcommand{\HII}{\ion{H}{2}}

\newcommand{\lzifu} {{\scshape lzifu}}
\newcommand{\ppxf} {{\scshape ppxf}}
\newcommand{\mpfit} {{\scshape mpfit}}
\newcommand{\pyneb} {{\scshape pyneb}}

\newcommand{\mappingsv} {{\scshape mappings~v}}
\newcommand{\mappings} {{\scshape mappings}}
\newcommand{\mappingsiii} {{\scshape mappings~iii}}
\newcommand{\hiiphot} {{\scshape hiiphot}}

\graphicspath{{./}{figures/}}

\received{??}
\revised{??}
\accepted{??}
\submitjournal{ApJL}

\shorttitle{Electron temperature variations across spiral arm}
\shortauthors{Ho et al.}

\begin{document}
\begin{CJK*}{UTF8}{bkai}

\title{Mapping electron temperature variations across a spiral arm in NGC\,1672}

\correspondingauthor{I-Ting Ho}
\email{iting@mpia.de}

\author[0000-0002-0757-9559]{I-Ting Ho (何宜庭)}
\affiliation{Max Planck Institut f\"{u}r Astronomie, K\"{o}nigstuhl 17, 69117 Heidelberg, Germany}

\author{Kathryn Kreckel}
\affiliation{Max Planck Institut f\"{u}r Astronomie, K\"{o}nigstuhl 17, 69117 Heidelberg, Germany}

\author{Sharon E. Meidt}
\affiliation{Sterrenkundig Observatorium, Universiteit Gent, Krijgslaan 281 S9, B-9000 Gent, Belgium}

\author{Brent Groves}
\affiliation{International Centre for Radio Astronomy Research, The University of Western Australia, 35 Stirling Hwy, 6009 Crawley, WA, Australia}
\affiliation{Research School of Astronomy and Astrophysics, Australian National University, Weston Creek 2611, Australia}

\author{Guillermo A. Blanc}
\affiliation{The Observatories of the Carnegie Institution for Science, 813 Santa Barbara Street, Pasadena, CA 91101, USA}
\affiliation{Departamento de Astronom\'ia, Universidad de Chile, Casilla 36-D, Santiago, Chile}

\author{Frank Bigiel}
\affiliation{ Argelander-Institut f\"{u}r Astronomie, Universit\"{a}t Bonn, Auf dem H\"{u}gel 71, 53121 Bonn, Germany}

\author{Daniel A. Dale}
\affiliation{Department of Physics \& Astronomy, University of Wyoming, Laramie, WY, USA}

\author{Eric Emsellem}
\affiliation{European Southern Observatory, Karl-Schwarzschild-Str. 2, 85748 Garching, Germany}
\affiliation{Universit\'{e} Lyon 1, ENS de Lyon, CNRS, Centre de Recherche Astrophysique de Lyon UMR5574, 69230 Saint-Genis-Laval, France}

\author{Simon~C.~O.~Glover}
\affiliation{Institut f\"{u}r theoretische Astrophysik, Zentrum f\"{u}r Astronomie der Universit\"{a}t Heidelberg, Albert-Ueberle Str. 2, D-69120 Heidelberg, Germany}

\author{Kathryn Grasha}
\affiliation{Research School of Astronomy and Astrophysics, Australian National University, Weston Creek 2611, Australia}

\author{Lisa J. Kewley}
\affiliation{Research School of Astronomy and Astrophysics, Australian National University, Weston Creek 2611, Australia}

\author{J.~M.~Diederik Kruijssen}
\affiliation{Astronomisches Rechen-Institut, Zentrum f{\"u}r Astronomie der Universit{\"a}t Heidelberg, M{\"o}nchhofstra{\ss}e 12-14, 69120 Heidelberg, Germany}

\author{Philipp Lang}
\affiliation{Max Planck Institut f\"{u}r Astronomie, K\"{o}nigstuhl 17, 69117 Heidelberg, Germany}

\author{Rebecca McElroy}
\affiliation{Max Planck Institut f\"{u}r Astronomie, K\"{o}nigstuhl 17, 69117 Heidelberg, Germany}

\author{Rolf-Peter Kudritzki}
\affiliation{Institute for Astronomy, University of Hawaii, 2680 Woodlawn Drive, Honolulu, HI 96822, USA}
\affiliation{University Observatory Munich, Scheinerstr. 1, D-81679 Munich, Germany}

\author{Patricia Sanchez-Blazquez}
\affiliation{Departamento de F\'{i}sica Te\'{o}rica, Universidad Aut\'{o}noma de Madrid, Cantoblanco, E-28049 Madrid, Spain}

\author{Karin Sandstrom}
\affiliation{Center for Astrophysics and Space Sciences, Department of Physics, University of California, San Diego, 9500 Gilman Drive, La Jolla, CA 92093, USA}

\author{Francesco Santoro}
\affiliation{Max Planck Institut f\"{u}r Astronomie, K\"{o}nigstuhl 17, 69117 Heidelberg, Germany}

\author{Eva Schinnerer}
\affiliation{Max Planck Institut f\"{u}r Astronomie, K\"{o}nigstuhl 17, 69117 Heidelberg, Germany}

\author{Andreas~Schruba}
\affiliation{Max-Planck-Institut f\"{u}r extraterrestrische Physik, Giessenbachstrasse 1, D-85748 Garching, Germany}



\begin{abstract}
We report one of the first extragalactic observations of electron temperature variations across a spiral arm. Using MUSE mosaic observations of the nearby galaxy NGC\,1672, we measure the \NII$\lambda5755$ auroral line in a sample of 80 \HII\ regions in the eastern spiral arm of NGC\,1672. We discover systematic temperature variations as a function of distance perpendicular to the spiral arm. The electron temperature is lowest on the spiral arm itself and highest on the downstream side. Photoionization models of different metallicity, pressure, and age of the ionizing source are explored to understand what properties of the interstellar medium drive the observed temperature variations. An azimuthally varying metallicity appears to be the most likely cause of the temperature variations. The electron temperature measurements solidify recent discoveries of azimuthal variations of oxygen abundance based on strong lines, and rule out the possibility that the abundance variations are artefacts of the strong-line calibrations. 
\end{abstract}



\section{Introduction} \label{sec:intro}

Since the first discovery of radial abundance gradients in the interstellar medium (ISM) of nearby galaxies almost half a century ago \citep{Searle:1971kx}, numerous observational and theoretical efforts have been made to robustly measure the gradient and understand its physical origin. It is now well-established that almost all nearby, normal star-forming disks exhibit negative radial gradients in their gas-phase oxygen abundance with virtually the same characteristic slope  \citep{Zaritsky:1994lr,Sanchez:2014fk,Ho:2015hl,Kaplan:2016aa}, except for the very inner part and outer part of the disks \citep{Belfiore:2017aa,Sanchez-Menguiano:2018aa,Lian:2018aa,Zinchenko:2019aa}. The gradient informs us about the inside-out formation history of galactic disks \citep{Prantzos:2000vn,Chiappini:1997fk,Chiappini:2001fk,Fu:2009vn,Pilkington:2012lr}. 

\begin{figure*}
\centering
\includegraphics[width=\textwidth]{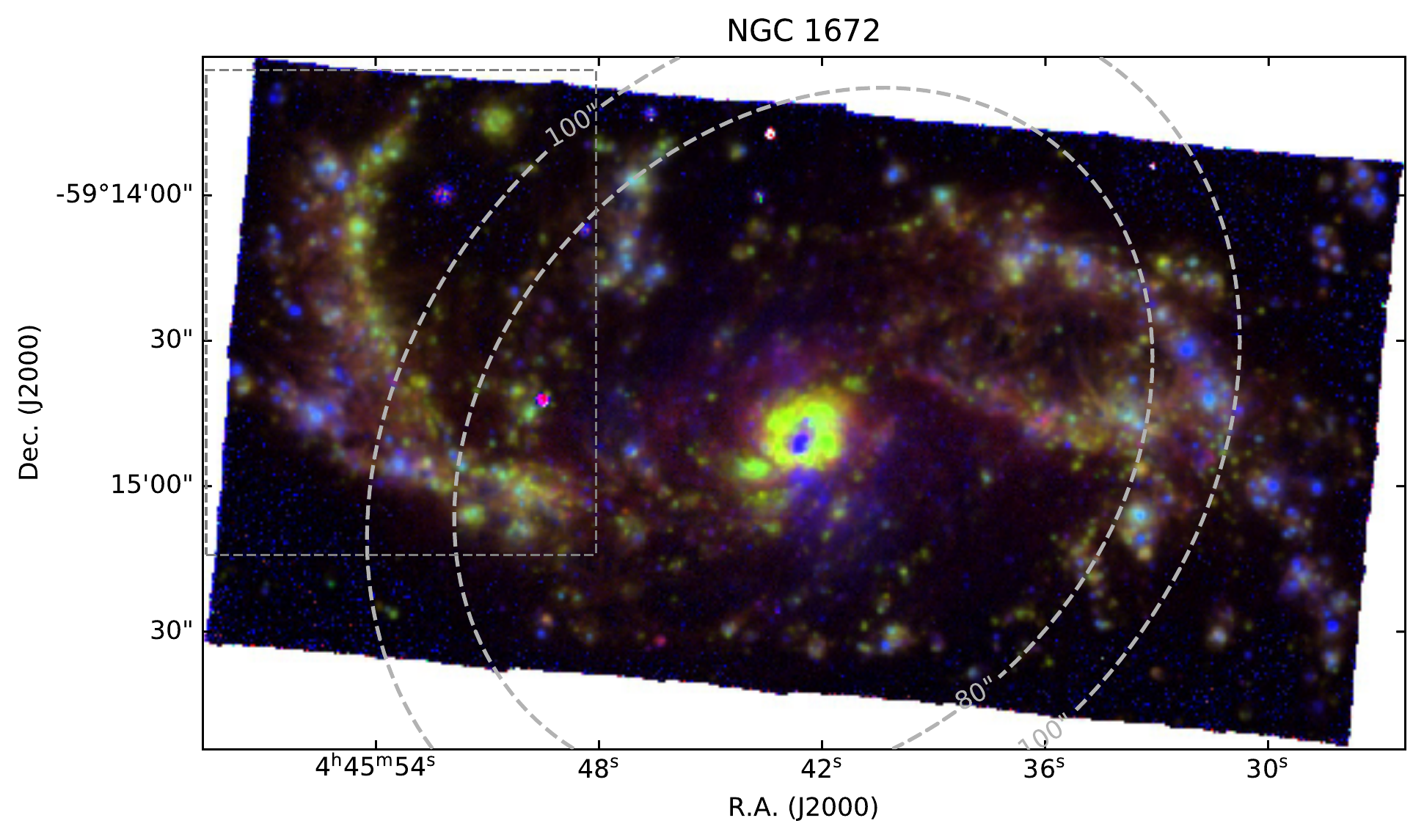}
\caption{
Three-color emission line image of NGC\,1672 from our MUSE mosaic observations at 0.8\arcsec\ (46~pc) resolution. Blue: [\ion{O}{3}]$\lambda$5007, green: H$\alpha$, red: [\ion{S}{2}]$\lambda$6716. 
Our analysis focuses on the eastern spiral arm segment inside the grey rectangle, the only location where we found azimuthal changes in the strong-line ratio. The two dashed contours indicate deprojected radii of 80\arcsec\ and 100\arcsec, discussed in Section~3. }\label{fig1}
\end{figure*}

Despite small dispersion around the linear gradients in typical isolated disks \citep[e.g.][]{Kennicutt:1996yf,Croxall:2016tg}, recent integral field spectroscopy observations have revealed systematic azimuthal variations of \HII\ region oxygen abundance in some galaxies. The reported variations are small, typically smaller than 0.1~dex and only in two cases up to 0.2~dex \citep{Vogt:2017aa,Ho:2017aa}. The variations usually correlate spatially with spiral structures \citep{Sanchez-Menguiano:2016zr,Kreckel:2019aa}, with the abundance typically reaching a maximum in the spiral arms and a minimum in the inter-arm regions \citep{Ho:2017aa,Ho:2018aa}. These observations imply that the dynamics of the density waves could affect the chemical enrichment of the ISM through possibly sub-kiloparsec-scale mixing, and open up the opportunity to compare with theoretical predictions for ISM mixing \citep{Roy:1995ys,de-Avillez:2002rw,Grand:2016hb,Ho:2017aa,Krumholz:2018aa}.

\begin{figure*}
\centering
\includegraphics[width=\textwidth]{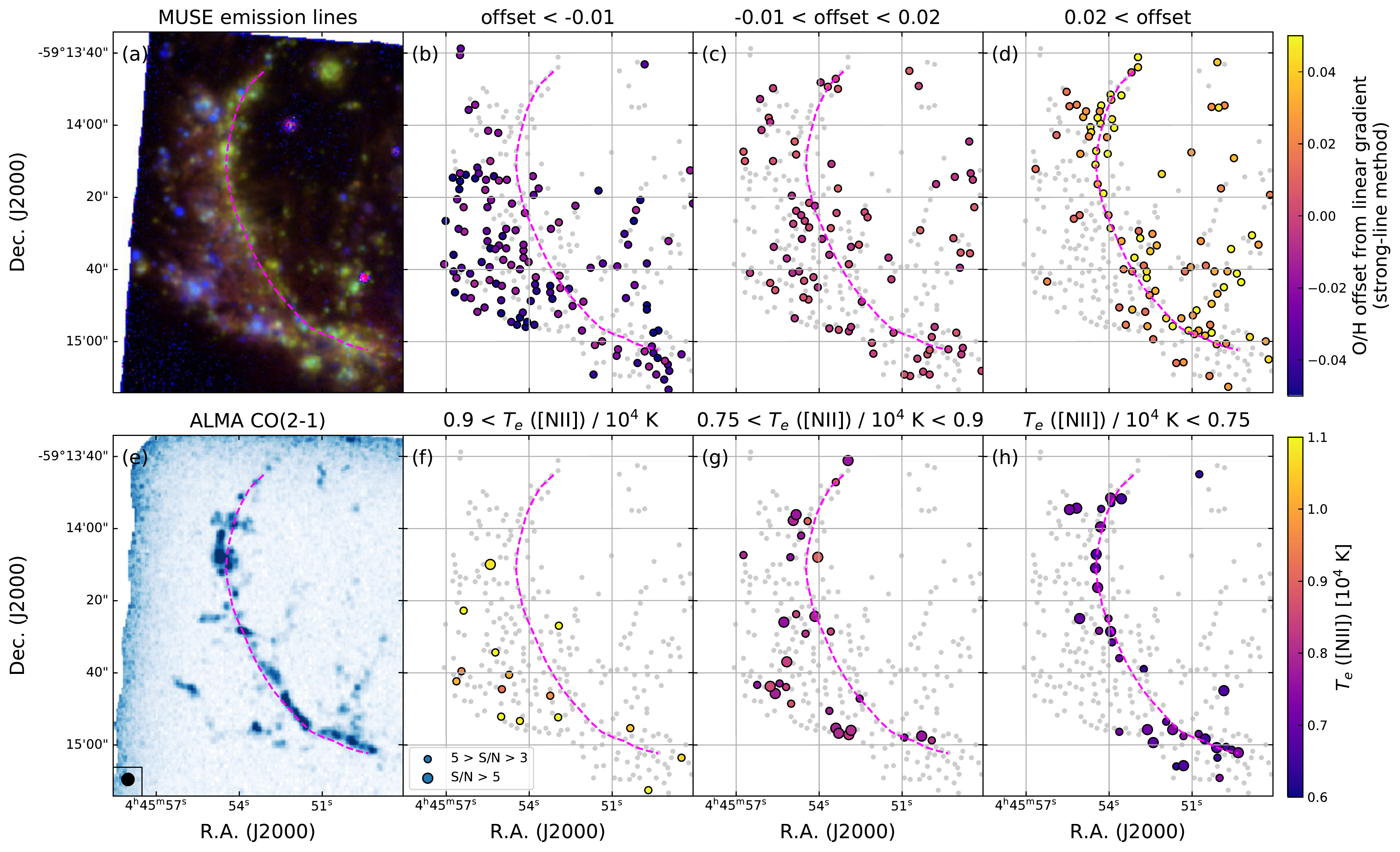}
\caption{ (a): Cutout of Figure~\ref{fig1} focusing on the eastern spiral arm. (b)--(d): Oxygen abundance offsets from the linear radial gradient \citep{Kreckel:2019aa}. The dots indicate positions of individual \HII\ regions. The abundances are derived using the S-cal strong-line method \citep{Pilyugin:2016aa}. The three panels correspond to offsets of different ranges, as indicated at the top of each panel. (e): ALMA CO(2-1) peak brightness map. (f)--(h): Electron temperature variations. The electron temperature inferred from the auroral line \NII$\lambda$5755 reflects the temperature of the low ionization zone. Typical $1\sigma$ error of the temperature is approximately 700~K for regions with $\rm 5>S/N > 3$. The ranges of $T_{\rm e}$ are indicated at the top of each panel. The magenta lines indicate the spiral arm location, defined by eye using panel (a).}\label{fig2}
\end{figure*}

So far, all extragalactic observations using integral field spectroscopy have relied on different strong-line calibrations to constrain the oxygen abundance in order to measure the azimuthal variations. However, all strong-line methods are subject to different degrees of systematic uncertainties that have not yet been fully understood and resolved \citep[e.g.][]{Kewley:2008qy,Peimbert:2017aa}. It remains unclear whether the varying strong-line ratios reflect changes of the actual gas-phase metallicity or other physical properties of the ISM. Understanding the physical mechanisms causing the line ratio variations is particularly important given that the observed variations are typically small (less than 0.1~dex). Other means of constraining the ISM metallicity using temperature sensitive lines (auroral or recombination lines; e.g.~\citealt{Li:2013nx}) are observationally very expensive. Here, we present one of the first extragalactic observations of azimuthal variations of electron temperatures ($T_{\rm e}$) in the eastern spiral arm of NGC~1672 (Figure~\ref{fig1}). By virtue of the excellent sensitivity of the Multi Unit Spectroscopic Explorer (MUSE) instrument on the Very Large Telescope (VLT), the \NII$\lambda5755$ auroral line is detected in a sample of 80 \HII\ regions. In a companion work, the strong-line oxygen abundances were also found to present systematic azimuthal variations (\citealt{Kreckel:2019aa}; panels (b) -- (d) in Figure~\ref{fig2}). Below, we will present the observational results and discuss what changes in the ISM could drive the systematic variations of $T_{\rm e}$.

Throughout the paper, we assume a distance of 11.9~Mpc, at which 1\arcsec\ is 57.7 parsec \citep{Tully:2009aa} and adopt an inclination angle of $37.5^\circ$ and a position angle of $135.7^\circ$ east of north, derived from CO rotation curve fitting (Lang et al. in preparation). 

\section{Observations, data reduction and data analysis}\label{sec:observations}

NGC\,1672 was observed as part of a large observing program with the MUSE spectrograph \citep{Bacon:2010ph} at the VLT (PHANGS\footnote{Physics at High Angular resolution in Nearby GalaxieS (https://www.phangs.org)}-MUSE survey; PI: Schinnerer). The inner spiral structure was mosaiced with eight pointings (Figure \ref{fig1}), each covering a 1\arcmin $\times$ 1\arcmin\ field of view with 0.2\arcsec\ pixels and a typical spectral resolution of $\sim$2.5\AA\ (FWHM) over the nominal wavelength range (4800--9300\AA).  Each pointing was observed in four rotations, with two sky pointings, and a total on source integration time of 43 minutes.  Typical seeing is 0.8\arcsec\ (or 46~pc). 

Details on the data reduction and \HII\ region catalog construction are contained in \citet{Kreckel:2019aa} and briefly summarized here. Observations are reduced using {\scshape pymusepipe}, a pipeline built on {\scshape esorex} and the MUSE pipeline recipes \citep{2016ascl.soft10004W} and developed by the PHANGS team\footnote{https://github.com/emsellem/pymusepipe}.  Absolute astrometric corrections for the MUSE pointings are derived from PHANGS team R-band imaging (as part of a narrow-band H$\alpha$ imaging campaign; details in Razza et al. in preparation), with astrometry determined from GAIA.  H$\alpha$ emission line maps are analyzed with \hiiphot\ \citep{Thilker:2000rf}, where `seeds' in the \HII\ region distribution are grown until they reach a termination criterion determined by the slope of the H$\alpha$ surface brightness.

Integrated spectra of \HII\ regions are then produced from the \HII\ region mask. The integrated spectra are then fit using \lzifu\ \citep{Ho:2016rc} that utilises \ppxf\ \citep{Cappellari:2011ys,Cappellari:2017aa} for modelling the underlying continuum and \mpfit\ \citep{Markwardt:2009lr} to fit the emission lines as single Gaussians. All the emission lines adopted in this study are fit simultaneously and share the same velocity and velocity dispersion. To reject regions with line ratios inconsistent with \HII-regions, the empirical line by \citet{Kauffmann:2003vn} is adopted for the \NII/H$\alpha$ versus \OIII/H$\beta$ diagram and the theoretical line by \citet{Kewley:2001lr} for the \SII/H$\alpha$ diagram. 

Electron temperature and density are derived from \NII$\lambda\lambda$5755,6584 and \SII$\lambda\lambda$6716,6731. The \pyneb\ module version 1.18 is adopted for extinction correction and solving iteratively for electron temperature and density using multi-level atomic models \citep{Luridiana:2012ve}. The default atomic data are adopted. We assume an intrinsic Balmer decrement (H$\alpha$/H$\beta$) of 2.85 and the CCM89 extinction curve \citep[][assume $R_V$ of 3.1]{Cardelli:1989qy}. The electron temperature derived from the \NII$\lambda\lambda$5755,6584 lines reflects the temperature of the low ionization zone of the nebula.

In total, eighty \HII\ regions with S/N of \NII$\lambda$5755 greater than 3 are analyzed and presented below (out of 225 detections). A subset of 33 regions has S/N greater than 5.

\begin{figure}
\centering
\includegraphics[width=\columnwidth]{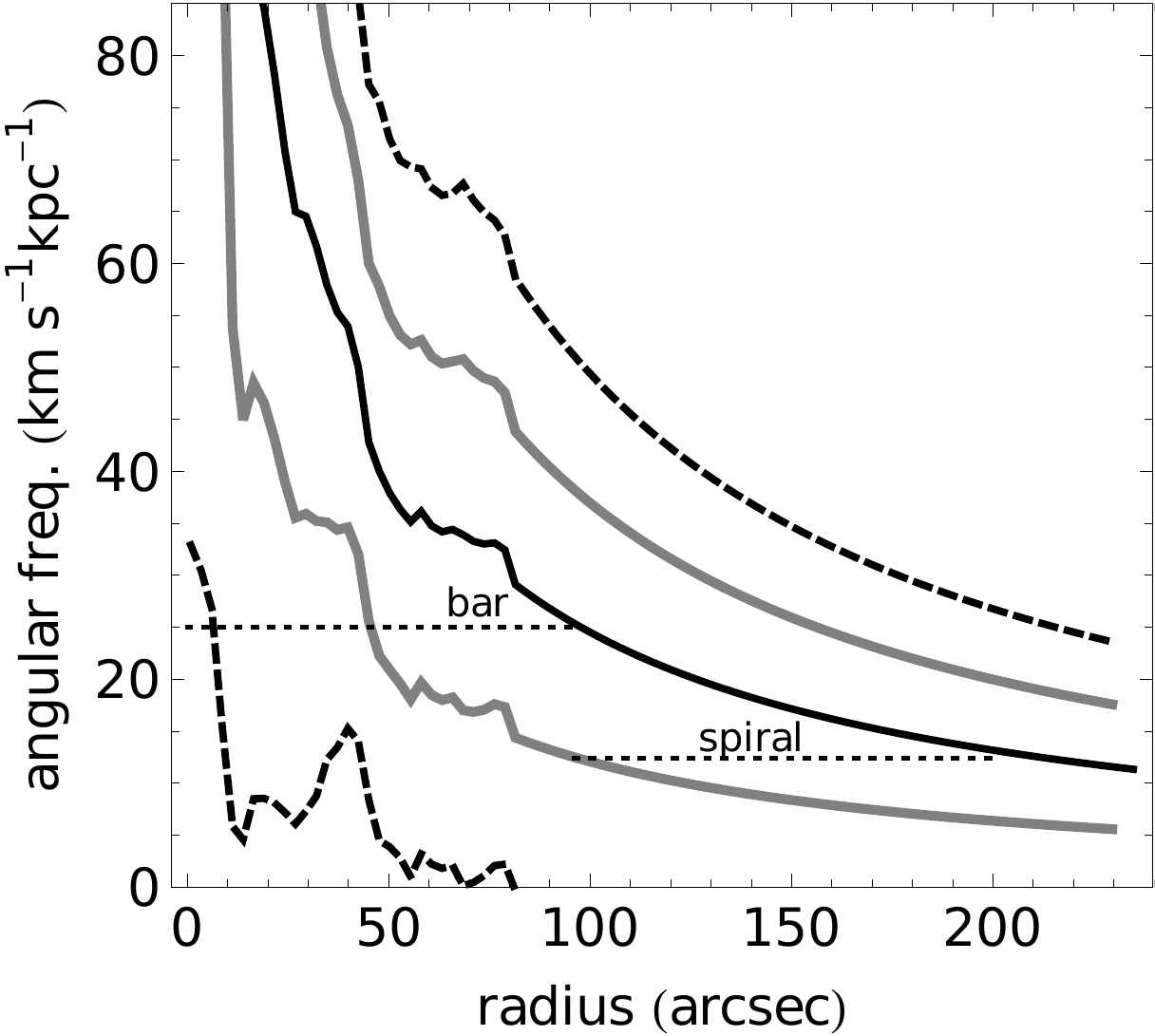}
\caption{Angular velocity curves in NGC 1672, $\Omega$ (solid), $\Omega\pm\kappa/2$ (Dashed) and $\Omega\pm\kappa/4$ (gray solid) derived from the observed rotation curve at $\rm R<80\arcsec$ measured by Lang et al. (in preparation) and extended to larger radius using the smooth model fitted by Lang et al. (in preparation). The adopted bar and spiral pattern speeds (see text) are marked by dotted horizontal lines. }\label{fig3}
\end{figure}

\begin{figure*}
\centering
\includegraphics[width=\textwidth]{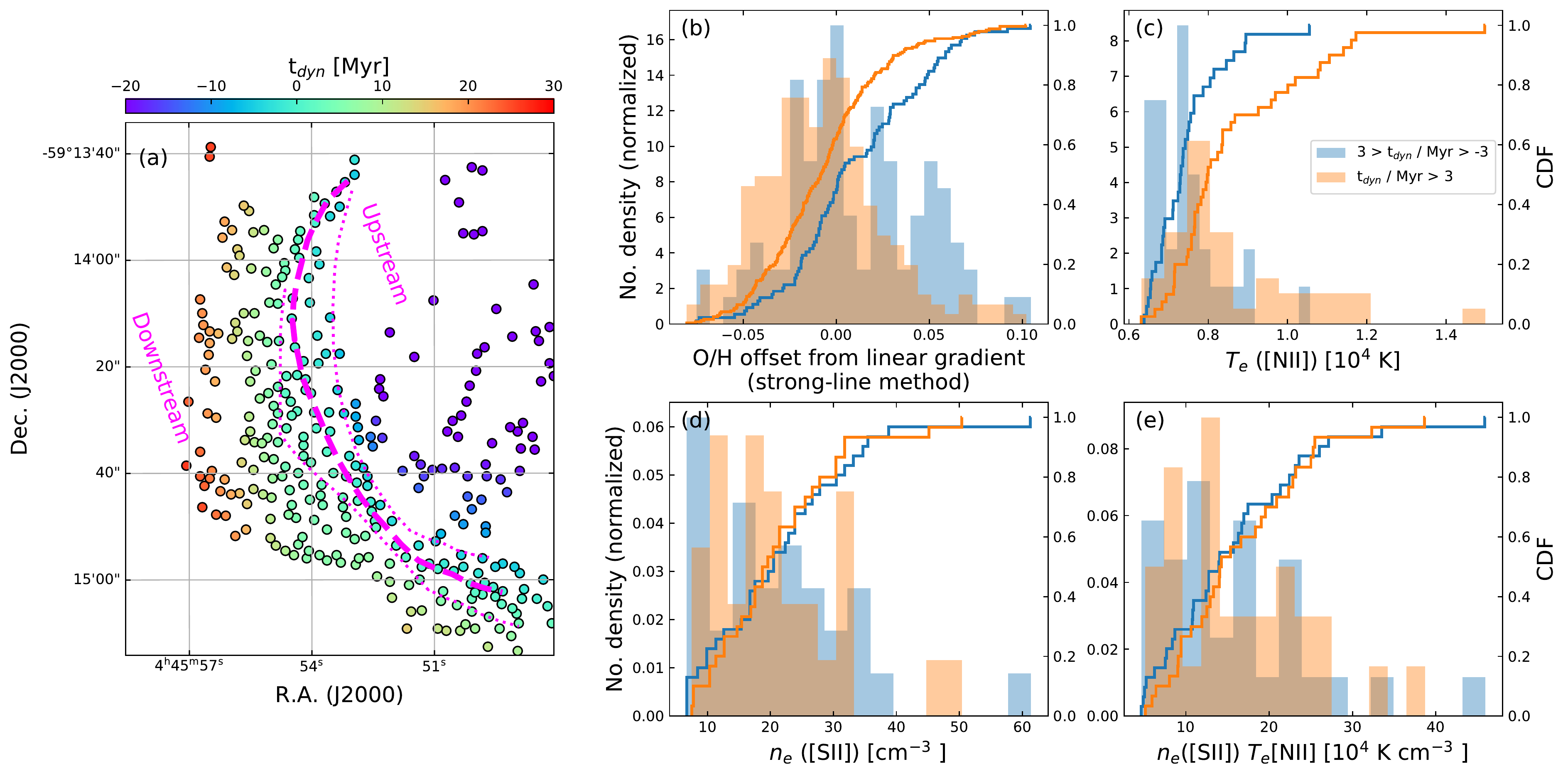}
\caption{(a): dynamical time since the spiral arm passage derived using our dynamical model (see Section~3). Positive time means the spiral arm has passed. The dashed magenta line indicates the spiral arm, and the two dotted lines $t_{\rm dyn}$ of +/-3~Myr. From (b) to (e): the two histograms represent regions of short ($3>t_{\rm dyn}>-3$~Myr) and long ($t_{\rm dyn}>3$~Myr) dynamical time, corresponding to spiral arm and downstream \HII\ regions. The curves indicate the corresponding cumulative distributions. The left $y$-axes correspond to the histograms and the right to the cumulative distributions. (b): strong line oxygen abundance offsets (from linear gradient; Figure~\ref{fig2} panels (b) to (d)). (c): electron temperature derived from the \NII\ auroral line. (d): electron density derived from the \SII\ doublet ratio. (e): the product of electron temperature and density as a tracer of pressure.
}\label{fig4}
\end{figure*}

\section{Results}

\subsection{Azimuthal electron temperature variations}
In panels (f)--(h) of Figure~\ref{fig2}, we present the electron temperature that is primarily constrained by the \NII\ auroral to strong line ratio.  To visualize the correlation with the spiral structure, we split the regions in three bins based on their temperatures. We define the spiral arm by eye guided by the colour figure in panel (a). The spiral arm defined on the optical emission line image (magenta curves) is very consistent with where bright molecular clouds live (panel (e), showing the ALMA CO(2-1) peak brightness map; \citealt{Sun:2018aa}; Leroy et al. in preparation). Figure~\ref{fig2} panels (f)--(h) show that the \HII\ regions with low temperatures ($T_{\rm e} < 7,500$K) preferentially coincide with the eastern spiral arm. Regions with higher temperatures are more likely to appear at the leading side of the spiral arm.

\subsection{Dynamical evolution and spiral pattern}

To further investigate possible correlations between ISM physical conditions and the spiral arm, we quantify the dynamical time since the spiral arm passage $t_{\rm dyn}=\pi/(\Omega-\Omega_p)$ where $\Omega_p$ is the pattern speed of the two-armed gas spiral and $\Omega=V_c/R$ is the angular (circular) rotation rate of material in the disk. This relies on a measure of the rotation curve $V_c$ and an estimate of the pattern speed of the dynamical structure in the disk.  We adopt the rotation curve measured from the CO velocity field out to 80\arcsec\ by Lang et al. (in preparation) and then extend this out to the edge of our field of view using the smooth rotation curve model fitted by Lang et al. (in prep.)

Our spiral pattern speed estimate is assigned in relation to that of the inner bar structure (as described below) and is chosen to yield a picture in which all \HII\ regions on the convex side of the arm lie inside the spiral corotation radius and thus downstream of the gas spiral.  Our estimate relies on first 
determining the inner bar pattern speed, which we assign based on the expectation that the bar ends at, or very near to, its corotation radius (\citealt{Contopoulos:1980aa} and see, e.g., \citealt{Elmegreen:1996aa}; \citealt{Rautiainen:2008aa}; \citealt{Corsini:2011aa}). We specifically assume $R_{CR}$=1.2 $a_B$ \citep{Athanassoula:1992aa}, where $a_B=80\arcsec$ is the bar length measured by \citet[][see also Figure~\ref{fig1}]{Herrera-Endoqui:2015aa}.  This yields $\Omega_{p,B} = 25~\rm km~s^{-1}~kpc^{-1}$.  We then estimate our lower spiral speed assuming that the bar and spiral arm are dynamically coupled (i.e.~\citealt{Masset:1997rz}; \citealt{Rautiainen:1999aa}; Figure~\ref{fig3}) and overlap at resonances defined by intersections with the angular frequency curves $\Omega-\kappa/m$ where $m$ is a positive integer and $\kappa$ is the radial epicyclic frequency in the disk.  Given the observed shape of the rotation curve, $\kappa\sim2\Omega$ over most of the disk, except towards the center.  This places the inner Lindblad resonance of the spiral (where $\Omega_p=\Omega-\kappa/2$) far from the bar corotation radius.  
We therefore assume that the bar corotation overlaps with the inner ultraharmonic resonance of the two-armed spiral (where $\Omega_p=\Omega-\kappa/4$), which is another possibility suggested by observations \citep{Meidt:2009aa}.  This yields the spiral pattern speed $\Omega_{p,S} = 12~\rm km~s^{-1}~kpc^{-1}$, placing all \HII\ regions inside the spiral corotation radius at approximately $200\arcsec$\footnote{Note that the scenario where the spiral rotates at the same speed as the bar is arguably ruled out by the presence of \HII\ regions on the convex side of the gas spiral arms, a scenario that requires that the regions to lie inside the spiral corotation radius.}. We adopt an uncertainty on the measured rotational velocities of $\pm$0.1 $V_c$ due to the inclination uncertainty (Lang et al., in preparation) and let this define the primary uncertainty on the measured timescale $t_{\rm dyn}$ at all positions, which corresponds to $\pm3$~Myr, on average across the radial zone of interest.

While a two pattern speed scenario is more realistic, we have also derived the dynamical time assuming only a single spiral pattern speed of $\Omega_{p,S} = 12~\rm km~s^{-1}~kpc^{-1}$. The conclusions of this work remain the same.

\subsection{Variations of ISM physical conditions}

In Figure~\ref{fig4}, we present the dynamical time ($t_{\rm dyn}$) in panel (a) and compare in panel (b) to (e) to the strong line metallicity offset, $T_{\rm e}$, $n_{\rm e}$ and pressure ($n_{\rm e} T_{\rm e}$). We split the \HII\ regions into two groups based on their dynamical time. The first group with $t_{\rm dyn}$ between $\pm3$~Myr\footnote{The negative dynamical time here is to take into account the width of the spiral arm and the uncertainty of defining the spiral arm by eye.} is directly associated with the spiral arm, whereas the second group, with $t_{\rm dyn}$ greater than 3~Myr, is not. Panels (d) to (e) compare the distributions of the two groups using both histogram and cumulative distribution function. Using two-side Kolmogorov--Smirnov test, we test the null hypothesis that the two distributions are drawn from the same distributions. The resulting {\it p-}values are $8.8\times10^{-4}$, $4.1\times10^{-4}$, 0.53, and 0.43 for panels (b), (c), (d), and (e), respectively. Figure~\ref{fig4} demonstrates that the spiral and downstream \HII\ regions, short and long dynamical time, respectively, are systematically different in electron temperature and strong line metallicity offset. There is no evidence that the electron density and pressure are systematically different.

We note that the product of $n_{\rm e}$ derived from \SII\ and $T_{\rm e}$ from \NII\ is only an approximation of the ISM pressure given that different ions are excited in different zones within an \HII\ region and that \HII\ regions may not be in pressure equilibrium with their surroundings. Deriving $n_{\rm e}$ from the \SII\ doublet also assumes that the temperature of the \SII\ and \NII\ zones are similar. Also, the \SII\ lines are not very sensitive to $n_{\rm e}$ in the low-density regime.

\section{Discussion}\label{sec:discussion}
\begin{figure*}
\centering
\includegraphics[width=\textwidth]{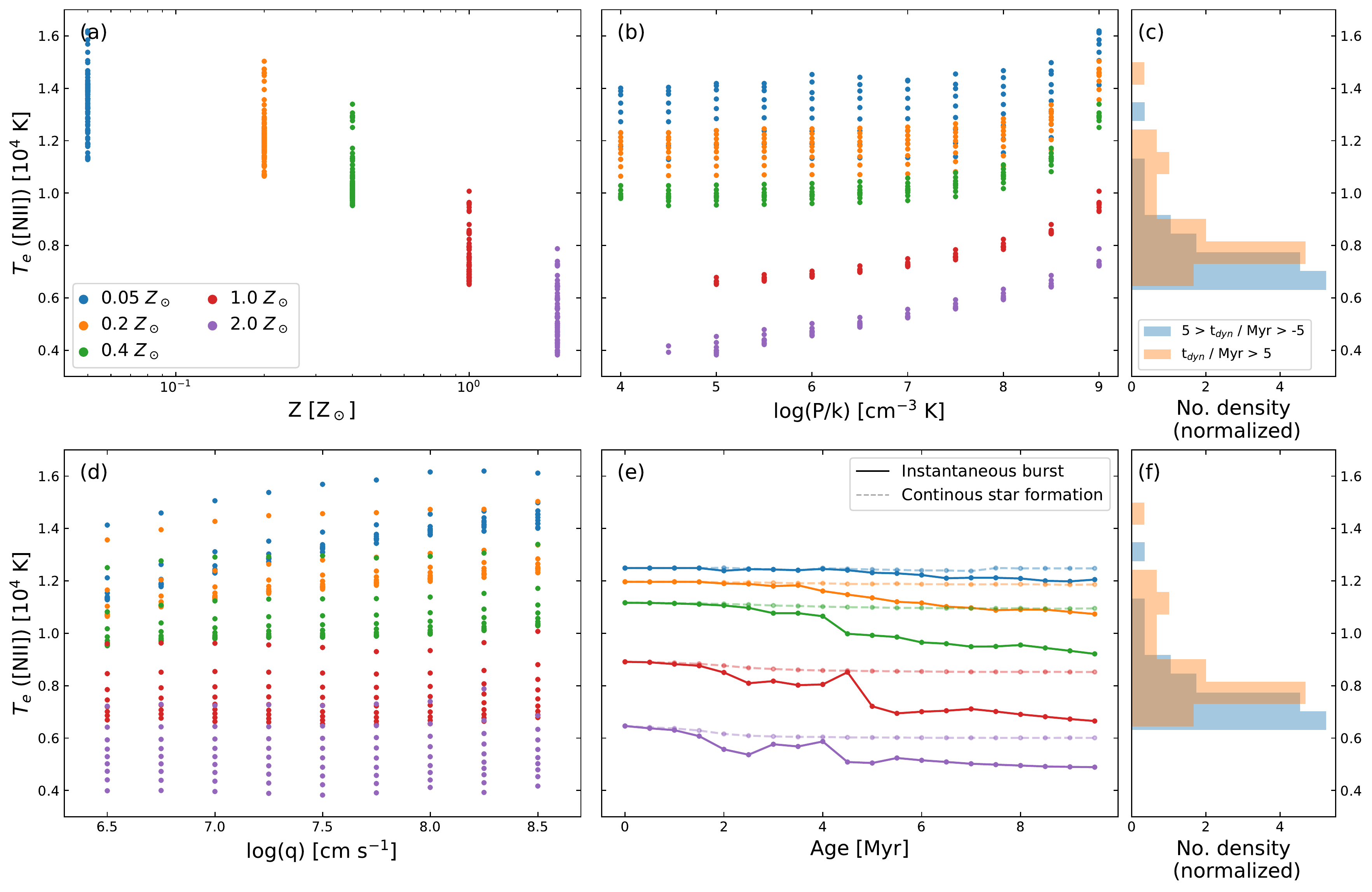}
\caption{(a), (b) and (d): $T_{\rm e}$(\NII) of \HII\ region models as functions of metallicity, pressure, and ionization parameter from the latest \mappingsv\ grids by \citet{Kewley:2019aa}. (e): $T_{\rm e}$(\NII) from the \mappingsiii\ models by \citet{Levesque:2010aa} as functions of ages and metallicities. These particular models have an input density of 100~$\rm cm^{-3}$ and ionization parameter of $\rm 2\times10^7~cm~s^{-1}$. The \mappings\ model line fluxes are treated like the observational data to derive $T_{\rm e}$(\NII) and $n_{\rm e}$(\SII) (Section~2). (c) and (f): Our \HII\ region temperature distribution for comparison. }\label{fig5}
\end{figure*}

\subsection{What drives the temperature variation?}

Our observations suggest that the \HII\ regions on the spiral arm have lower \NII\ temperature than the downstream side. There are several possibilities for this to occur. One of the most straightforward explanations is that the lower temperature on the spiral arm is due to the higher metal abundance that causes effective cooling of the nebula. However, at the spatial scale of marginally resolving individual \HII\ regions and dynamical timescales of a few Myr, variations in \NII\ temperature could also reflect changes in other ISM physical conditions or stellar characteristics.

The \NII\ temperature in an \HII\ region can be changed by the pressure of the nebula. In the latest \mappingsv\ photoionization models by \citet{Kewley:2019aa}, the authors construct model grids as functions of metal abundance, pressure, and ionization parameter. At solar and super-solar metallicities of their plane-parallel models, pressure correlates positively with \NII\ temperature (panel (b) of Figure~\ref{fig5}). From $\log(P/k)$ of 5 to 6 $\rm cm^{-3}~K$, the \NII\ temperature increases from approximately 4000 to 5000~K for twice-solar metallicity models. At half-solar metallicity, the positive correlation only starts at extreme pressure of $\log(P/k) > 7~\rm cm^{-3}~K$. Our \HII\ regions are at approximately 0.8 solar metallicity and $\log(P/k)=5~{\rm cm^{-3}~K}$, it is thus unlikely that the pressure can substantially change the \NII\ temperature. Moreover, if one attributes the temperature variation purely due to pressure variation, this would imply that the spiral arm \HII\ regions have lower pressure than those in the inter-arm regions. This is not supported by the data ($n_{\rm e} T_{\rm e}$; panel (e) of Figure~\ref{fig4}) and also contradicts the expectation of higher pressure in the spiral arm due to higher gas and star formation rate surface densities.

Another possible driver of the temperature variations is the aging of the ionizing source. As a stellar population ages, evolution of the spectral energy distribution alters the ionization structure of the \HII\ region and hence the measured electron temperature. To understand this effect, we derive \NII\ temperatures at different ages from the \mappingsiii\ models by \citet[][see panel(e) of Figure~\ref{fig5}]{Levesque:2010aa}. For the instantaneous burst models\footnote{The instantaneous burst models are more appropriate than the continuous star formation history for comparing with our data given that our spatial resolution can almost isolate individual \HII\ regions.}, a systematic decrease in \NII\ temperature of up to 2,000~K is apparent for the first 10~Myr of the stellar evolution for the models with metallicities above 0.4 solar. If star formation occurs only on the spiral arm, then as stars age and drift off the spiral arm, we would expect the temperature of the \HII\ regions to decrease, leading to lower temperatures on the downstream side. This is opposite to the observation where the \HII\ regions on the spiral arm have lower temperatures than those on the downstream side. The reality is likely more complex given that the dynamical time of some of our regions is much longer than the typical lifetime of \HII\ regions (up to 30~Myr versus a few Myr), which suggests that many downstream \HII\ regions are inconsistent with being formed on the spiral arm. It is unlikely that the aging effect alone could produce the observed temperature variations. 

\subsection{Azimuthal variations of metallicity}

Of the three possibilities discussed above, the most likely cause of the observed temperature variations is the higher ISM metallicity on the spiral arm than the downstream side. The $T_{\rm e}$ measurements would thus confirm the finding from the strong-line method (panels (b)--(d) in Figure~\ref{fig2}). That is, the inverse correlation between strong-line metallicitiy and $T_e$ reflects simply metal cooling of the nebula.

To quantify the absolute abundance variation from the $T_{\rm e}$ measurements is non-trivial. The \OIII$\lambda4363$ auroral line and the \OII$\lambda\lambda3726,3729$ strong lines fall outside the MUSE passband, so it is not possible to directly constrain the oxygen abundance. Alternatively, we can compare the observationally-derived $T_{\rm e}$ to the \mappingsv\ models by \citet{Kewley:2019aa}. The median $T_{\rm e}$ is 7294~K for the spiral arm regions and 7980~K for the downstream regions (Figure~\ref{fig4} panel(c)). This corresponds to 0.81 and 0.77 solar metallicity\footnote{We linearly interpolate between the 0.5 and 1 solar metallicity models}, respectively, using the models of $\log(P/k)$ of $\rm5~ cm^{-3}~K$ and ionization parameter of $10^7\rm~cm~s^{-1}$. The abundance difference of 0.04~dex is larger than the 0.01 to 0.02~dex from the strong line method (Figure~\ref{fig4} panel(b)). However, given the calibration uncertainty of the strong line method, particularly for high-metallicity regions, and the simple geometry and sparse metallicity sampling of the models, we consider that the two estimates are not inconsistent.

Despite that a quantitative statement to an accuracy of a few times 0.01~dex is not yet feasible, it is clear that qualitatively \HII\ regions on the spiral arm have higher oxygen abundance than those at the downstream side. Such a signature has been reported in a number of nearby spiral galaxies using different strong-line methods (\citealt{Sanchez-Menguiano:2016zr,Vogt:2017aa,Ho:2017aa,Ho:2018aa}). In \citet{Ho:2017aa}, the authors argue that the decrease in metallicity is due to mixing triggered by a spiral density wave. Prior to density wave passage, chemical inhomogeneity (between star-forming and non star-forming sites) was established due to inefficient mixing, when materials travel in the inter-arm region. The passage of the spiral density wave drives large-scale mixing with the surrounding lower metallicity gas, resulting in the low metallicity and high temperature in the downstream \HII\ regions. It remains unclear why such a signature is only observed in some galaxies, and even more puzzling why sometimes only seen in one spiral arm (e.g.\ this study; also see \citealt{Kreckel:2019aa}). It could be that right combinations of dynamical time, star formation and gas surface densities are required to drive large metallicity variations. Dynamical properties of spiral arms may also be important (e.g. short- versus long-lived arms, material arm versus density wave), and the mixing processes could be more complicated than that proposed in \citet{Ho:2017aa}. Simulations of the dynamical effect of spiral arms on star formation and chemical mixing, and also observations of multiple temperature sensitive lines (e.g.~the CHAOS Project; \citealt{Berg:2015aa,Croxall:2015aa,Croxall:2016tg}) and resolved stellar populations with the Hubble Space Telescope could be possible ways forward.

\acknowledgments

This work is based on observations collected at the European Organisation for Astronomical Research in the Southern Hemisphere under ESO programme 1100.B-0651.
JMDK gratefully acknowledge funding from the German Research Foundation (DFG) in the form of an Emmy Noether Research Group (grant number KR4801/1-1) and the DFG Sachbeihilfe (grant number KR4801/2-1). JMDK gratefully acknowledges funding from the European Research Council (ERC) under the European Union's Horizon 2020 research and innovation programme via the ERC Starting Grant MUSTANG (grant agreement number 714907). F.B. acknowledges funding from the European Union’s Horizon 2020 research and innovation programme (grant agreement No 726384).

\bibliography{references}


\end{CJK*}
\end{document}